\documentclass[aps,prl,twocolumn]{revtex4-1}

\usepackage{mathtools}
\usepackage{amssymb}
\usepackage{graphicx}
\usepackage{bbold}
\usepackage{hyperref}
\usepackage{xcolor}
\usepackage{enumerate}

\hypersetup{colorlinks=true, linkcolor=blue, urlcolor=blue, citecolor=blue}

\begin{document}

\title{Power-efficiency-stability trade-off in quantum information engines}

\author{Milton Aguilar}
	\email{maguilar@itp1.uni-stuttgart.de}
	\affiliation{Institute for Theoretical Physics I, University of Stuttgart, D-70550 Stuttgart, Germany}
	
\author{C\"uneyt \"Unal}
	\affiliation{Institute for Theoretical Physics I, University of Stuttgart, D-70550 Stuttgart, Germany}

\author{Eric Lutz}
	\affiliation{Institute for Theoretical Physics I, University of Stuttgart, D-70550 Stuttgart, Germany}


\begin{abstract}
Efficiency and power are two central measures of the performance of  thermal machines. We here study the power-efficiency-stability trade-off in a finite-time quantum Carnot information engine, in which an information reservoir replaces the usual cold bath of a quantum Carnot engine. We analytically evaluate mean and variance of the work output, and demonstrate that maximum efficiency can be reached at both finite work output and finite work output fluctuations. We additionally show that the relative work output fluctuations may be smaller than those of the corresponding Carnot heat engine. This result implies that the finite-time quantum Carnot information engine can be more stable than the quantum Carnot heat engine, an important property for practical applications.
\end{abstract}

\maketitle

Efficiency and power are two essential figures of merit of thermal engines. Efficiency, defined as the ratio of work output and heat input, provides a measure of heat-to-work conversion, while power characterizes the work output  rate \cite{cen01}. The second law limits the efficiency of any heat engine operating between thermal baths to be smaller than the Carnot efficiency, $\eta_\text{C}= 1-T_{\text{c}}/T_{\text{h}}$, where $T_{\text{c,h}}$ are the  cold and hot bath  temperatures  \cite{cen01}. By contrast, thermodynamics does not constrain the power output. It is commonly assumed that there is a trade-off between the two quantities, since efficiency is maximum  in the quasistatic regime where  power vanishes, and vice versa \cite{whi14,pro16,raz16,shi16,pol17}. The only  way to increase power would accordingly  be to sacrifice efficiency \cite{che94,bro05,sch07,esp09,esp10}. However, this premise has recently been challenged in a growing number of studies that have shown that maximum efficiency may be approached at nonzero power \cite{hol14,pol15,cam16,hol17,pie18,lia25,hol18,miu22,ber22,vu24}, with either divergent \cite{hol14,pol15,cam16,hol17,pie18,lia25} or finite \cite{hol18,miu22,ber22,vu24} power fluctuations. These findings suggest that it is possible to realize useful machines that run at optimal efficiency.

We here address the fundamental question of the attainability of maximum efficiency at finite power   in quantum information engines. Information engines extract work from a single heat reservoir by exploiting information gained about  the working medium, for example, with the help of   cyclic measurement and feedback operations \cite{cao09,sag10,abr11,hor11,bau12,sag12,esp12,man13,hor13,um15,par16,yam16,hor19}. They are inherently different from heat engines, since the second law prevents cyclic work extraction from a single heat bath  \cite{cen01}. They are made possible thanks to  the intimate link between information  and thermodynamic entropies \cite{mar09,par15,lut15}. Information engines may be regarded as interacting with one heat reservoir and one information reservoir that  only exchanges entropy, but no energy, with the device \cite{def13,bar14,bar14a}. Successful information-to-work conversion has been reported in numerous classical  \cite{toy10,rol14,kos14,kos14a,kos15,vid16,chi17,rib19,pru25,pan18,adm18,pan20,sah21,sah23,yan24} and quantum  \cite{cam16a,cot17,mas18,nag18,naj20} experiments.

\begin{figure}[t]
	\centering
	\includegraphics[width=.45\textwidth]{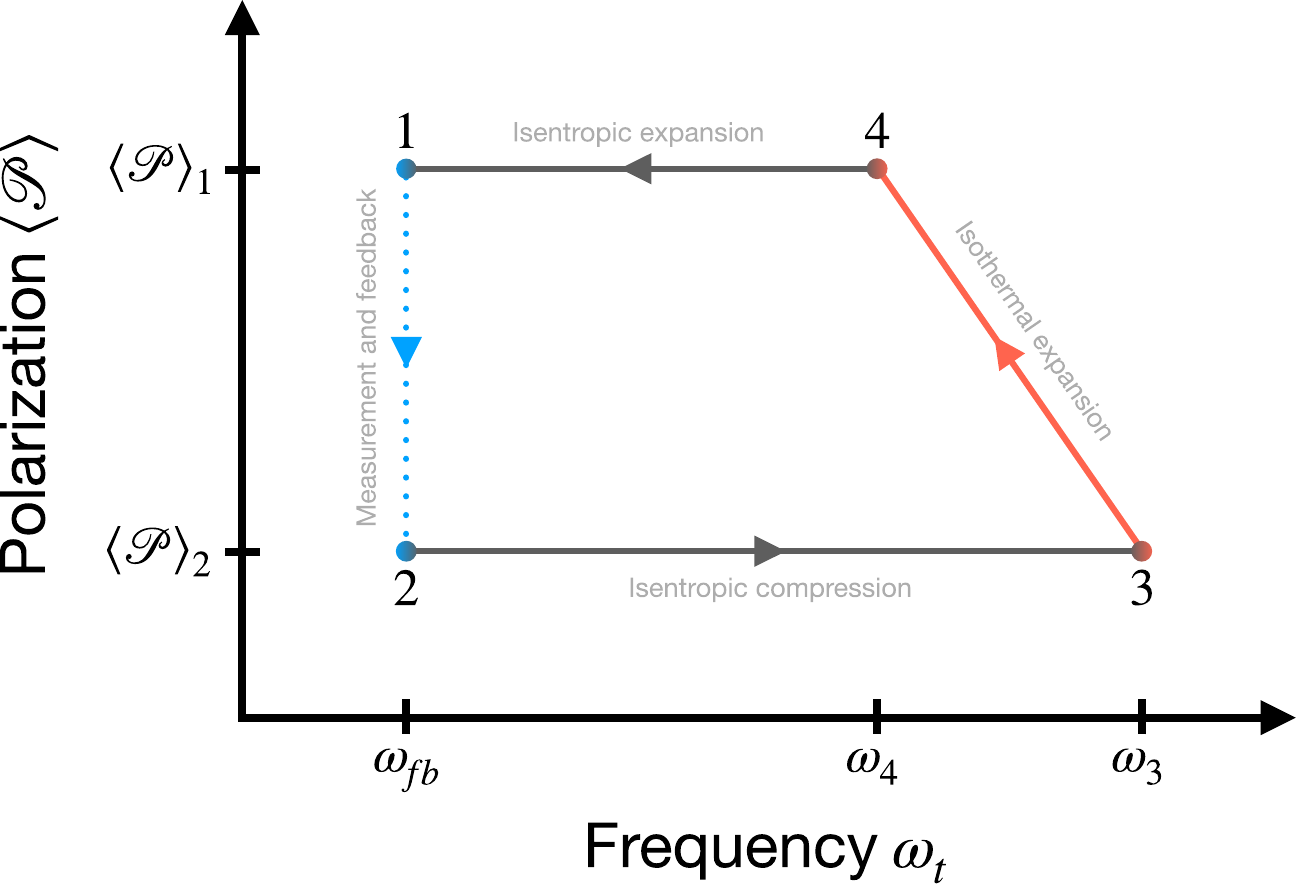}
	\caption{Finite-time quantum Carnot information cycle. The polarization-frequency diagram  for a working medium with Hamiltonian ${H}_{\text{t}} = \omega_{\text{t}} {\mathcal{P}}$, with frequency $\omega_{\text{t}} $ and generalized polarization ${\mathcal{P}}$, begins with a measurement-and-feedback operation (1-2) that cools the system to a fixed thermal state. The remainder of the cycle is that of a standard finite-time quantum Carnot heat engine which consists of an isentropic compression (2-3) followed by an isothermal expansion (3-4) and an isentropic expansion (4-1) back to the initial state.} 
        \label{fig:cycle}
 \end{figure}

In the following, we investigate the power-efficiency-stability trade-off in a finite-time quantum Carnot information engine, in which an information reservoir replaces the usual cold bath of a finite-time quantum Carnot engine (Fig.~\ref{fig:cycle}) \cite{fal23}. The reversible Carnot cycle describes the most efficient heat engine, and  hence plays  a special role  in thermodynamics \cite{cen01}. Similarly, the quantum Carnot information engine is fully reversible,  and reaches maximum efficiency,  in the limit of long cycle times, like its heat engine counterpart \cite{qua07,abe11,dan20,abi20,den21}. We  analytically evaluate mean and variance of the work output for a generic working medium, and demonstrate that maximum efficiency  can be reached at both finite work output and finite work output fluctuations. We moreover show that the relative work output  fluctuations may be smaller than those of the corresponding Carnot heat engine \cite{den21}. This result  reveals that the finite-time  Carnot information engine can be more stable than the associated  heat engine, an interesting feature for concrete applications.

\paragraph{Quantum Carnot information engine.} We consider a finite-time quantum Carnot information engine that consists of two isentropic (expansion and compression) branches, as well as one (hot) isothermal compression step and one reversible measurement-plus-feedback operation (Fig.~\ref{fig:cycle}) \cite{fal23}. The latter step replaces the usual (cold) isothermal expansion of the finite-time quantum Carnot cycle \cite{qua07,abe11,dan20,abi20,den21}. In order to make the {isentropic} expansion and compression branches reversible, and avoid quantum friction \cite{kos02,kos03,zam14}, the Hamiltonian is chosen to  commute 
with itself at all times, $[H_\text{t} , H_{\text{t'}} ] = 0$, as in the standard quantum Carnot cycle \cite{qua07,abe11,dan20,abi20,den21}. For concreteness, and without
loss of generality, we assume that the working medium is described by  a Hamiltonian of the
scaling form $H_{\text{t}} = \omega_{\text{t}} {\cal P}$, with time-dependent frequency $\omega_{\text{t}}$ \cite{den21}; for a two-level system, ${\cal P}$ is   the polarization. We further denote the 
 instantaneous thermal state, at temperature $T$,  as $\rho (\omega_{\text{t}} , T) = \text{exp} (- \omega_{\text{t}} \mathcal{P} / T) / Z$ with partition function $Z = \text{tr} [ \text{exp} (- \omega_{\text{t}} \mathcal{P} / T) ]$.
 
  The four branches of the quantum information cycle may then be characterized as follows (Fig.~\ref{fig:cycle}) \cite{fal23}:  (1-2) the  reversible measurement-and-feedback  process transforms any input thermal state $\rho (\omega_{\text{fb}} , T_1)$ into the fixed output thermal state  $\rho (\omega_{\text{fb}} , T_2)$, with $T_{{2}} < T_{{1}}$.
  The measurement is specifically implemented using a generalized measurement  with positive operators {$ \{M_{i}^{\dagger} M_{i}\}$} that satisfy $\sum_i  M_i^\dagger M_i =I$ \cite{jac14}. The  state after the measurement is accordingly given by  $\rho_{i} = M_{i} \rho (\omega_{\text{fb}} , T_{{1}}) M_{i}^{\dagger} / p_i$ with probability $p_{i} = \text{tr} [ M_{i} \rho (\omega_{\text{fb}} , T_{{1}}) M_{i}^{\dagger} ]$. In order to ensure thermodynamic reversibility, the measurement operators are  chosen such that $[ M_{i} , \rho (\omega_{\text{fb}} , T_{{1}}) ] = 0$ \cite{jac14}, implying that they describe  nonprojective energy measurements. Depending on the measurement outcome, a feedback controller transforms the state $\rho_{i}$ into $\rho (\omega_{\text{fb}} , T_{{2}})$ while preserving the von Neumann entropy,  $S(\rho_{i}) = S(\rho(\omega_{\text{fb}} , T_{{2}})) $, for all $i$ \cite{fal23}. (2-3) An isentropic compression unitarily maps the state $\rho (\omega_{\text{fb}} , T_{{2}})$ onto $\rho (\omega_{3} , T_{\text{h}})$ by increasing the frequency to $\omega_{3} > \omega_{\text{fb}}$, and increasing the temperature  to that of  the hot bath $T_{\text{h}}$. (3-4) An isothermal expansion additionally  brings the state $\rho (\omega_{3} , T_{\text{h}})$ to $\rho (\omega_{4} , T_{\text{h}})$ by decreasing the frequency  to $\omega_{4} < \omega_{3}$, while the system is weakly coupled to the hot bath. (4-1) Finally, the cycle is closed with an isentropic expansion that unitarily brings the state $\rho (\omega_{4} , T_{\text{h}})$  back to $\rho (\omega_{\text{fb}} , T_{{1}})$. 
 
 Since the  two {branches} are reversible irrespective of their duration, it is customary to set the latter times to zero for simplicity \cite{gev92,lin03}. The cycle time is accordingly given by the sum of  the durations of the measurement-and-feedback protocol and of the isothermal step, $\tau= \tau_{\text{fb}} + \tau_{\text{h}}$. The total work output  is, moreover,  the sum of the work produced by the working medium  and of the work extracted by the measurement-and-feedback protocol, $W=W_{\text{wm}}+W_{\text{fb}}$. It has been found to be equal to the heat $Q_{\text{h}}$ exchanged during the hot isothermal compression \cite{fal23}. In the long-time limit, we therefore have to lowest order, $W=Q_{\text{h}} = T_{\text{h}} ( \Delta S - \Sigma_{\text{h}} / \tau_{\text{h}} )$ (corresponding to the low dissipation regime \cite{esp10}), where $\Delta S$ is the von Neumann entropy change of the working substance, $T_{\text{h}}$ is the temperature of the (hot) bath, and $\Sigma_{\text{h}}$ is a coefficient that characterizes the entropy production during the time $\tau_{\text{h}}$ it takes the isothermal expansion to be completed. The power then follows as $\mathcal{P} = |W|/\tau$.  On the other hand, the  efficiency at which information is converted into work in the cyclic  information engine is defined as $\eta ={ {|W|}}/{T_{\text{h}} \Delta S} = 1 - {\Sigma_{\text{h}}}/{\Delta S \tau_{\text{h}}}$
\cite{qua06,jac09,kim11,str13,bra15,elo17,elo18,sea20}.

\begin{figure*}[t]
	\centering
	\includegraphics[width=1\textwidth]{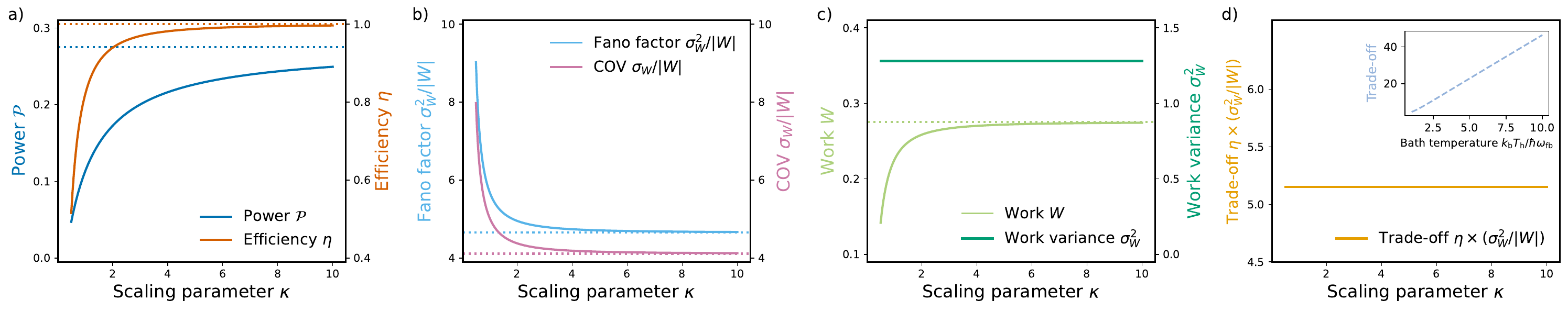}
	\caption{Power-efficiency-stability trade-off of a two-level Carnot information engine. In the limit of fast thermalization, characterized by the scaling parameter $\kappa$, (a) both efficiency and power increase, while (b) the relative work fluctuations, quantified by the Fano factor and the coefficient of variation (COV), decrease. In this regime, the machine achieves maximum efficiency at finite power and small power fluctuations. (c) The average work output grows steadily, whereas the corresponding variance remains constant (the work done during the isothermal expansion is deterministic). (d) The product of efficiency and Fano factor is a constant that increases with the hot bath temperature $T_{\text{h}}$ (Inset: the trade-off is shown as a function of the dimensionless hot bath temperature $k_{\text{b}} T_{\text{h}} / \hbar \omega_{\text{fb}}$). Dotted lines show the asymptotic expressions \eqref{4}. Parameters  are $\omega_{\text{fb}} = 1$, $\omega_{3} = 2 \omega_{4} = 4$, $\tau_{\text{h}} = \tau_{\text{fb}} = 1$, $\gamma = 2 \alpha = 2$, and $T_{\text{h}} = 1$.} 
        \label{fig:scaling}
 \end{figure*}
 
\paragraph{Work output fluctuations.} We next compute the variance, $\sigma_{w}^{2} = \langle (w - W)^{2} \rangle$, of the stochastic work output $w$. To that end, we  evaluate the work output  distribution $P (w)$. As before, we decompose the total stochastic work as  $w = w_{\text{wm}} + w_{\text{fb}}$, and compute the corresponding distributions, $P_{\text{wm}}(w_{\text{wm}})$ and $P_{\text{fb}}(w_{\text{fb}})$, separately.  Since the input and output thermal  states, $\rho (\omega_{\text{fb}} , T_{{1}})$ and $\rho (\omega_{\text{fb}} , T_{{2}})$,  of the measurement-and-feedback branch (1-2) only depend on the  frequency  $\omega_{\text{fb}}$ and on the temperatures $T_{{1}}$ and $T_{{2}}$, but not on the details of the  other branches   of the cycle, $w_{\text{wm}}$ and $w_{\text{fb}}$ are independent variables. As a result,  the total work distribution is the convolution of the respective probability distributions, $P(w) = P_{\text{wm}} \ast P_{\text{fb}} (w)$ \cite{pap91}. A standard procedure to determine this distribution is the two-point-measurement scheme \cite{tal07,esp09a,cam11}, where projective energy measurements are performed at the four corners of the cycle \cite{den20,jia21,dar21,den24}. However, this method  is not suitable here, since   projective measurements   perturb the engine, in particular, the fine-tuned measurement-and-feedback step.  {For this reason, we employ the end-point measurement scheme}  \cite{ghe21,gia23,her23}, where two copies of the system are analyzed instead of one: one of the copies is measured before the process takes place, and the second copy is left to evolve undisturbed and measured at the end. We then obtain (Supplemental Material)
\begin{eqnarray}
	P (w)  & =& \sum_{j,k,l,m} \sum_{p,q,r,s}  p_{3}^{k} p_{2}^{j} p_{1}^{m} p_{4}^{l} p_{1}^{q}  p_{1}^{p} p_{2}^{s}  p_{1}^{r} \nonumber \\
	& \times& \delta \big [ w + (E_{3}^{k} - E_{2}^{j}) - W_{\text{h}} +  (E_{1}^{m} - E_{4}^{l}) \nonumber \\
	& +& (E_{1}^{q} - E_{1}^{p}) + (E_{2}^{s} - E_{1}^{r}) \big],
	\label{eq:pFInal} 
\end{eqnarray}
where $E_z^i$ is an eigenvalue of the Hamiltonian $H_z$ of the working medium at point $ z= (1, 2, 3, 4)$ in the cycle (Fig.~\ref{fig:cycle}), and $p_z^i$ is the  probability to projectively measure that eigenvalue. Furthermore, $W_\text{h} = Q_{\text{h}} - \Delta U_{\text{h}}$ is the work done during the isothermal expansion (3-4), with the associated change of  internal energy  $\Delta U_{\text{h}} = \text{tr} [ \rho (\omega_{4} , T_{\text{h}}) {H}_{4} ] - \text{tr} [ \rho (\omega_{3} , T_{\text{h}}) {H}_{3} ]$. One may  verify that the mean work  is equal to $W = \langle w \rangle = \{ \text{tr} [ \rho (\omega_{4} , T_{\text{h}}) {H}_{4} ] - \text{tr} [ \rho (\omega_{3} , T_{\text{h}}) {H}_{3} ] \} + W_\text{h} = Q_{\text{h}}$, as it should. We stress that the last equality would have been violated using the two-point-measurement scheme \cite{tal07,esp09a,cam11} or dynamic Bayesian networks \cite{nea03,dar09,mic20,par20,mic24} (Supplemental Material).

The variance of the work  of the finite-time quantum Carnot information engine follows from Eq.~\eqref{eq:pFInal} as 
\begin{equation}
	\sigma_{w}^{2} = T_{\text{h}}^{2} \left[ f_{5} (T_{{1}})  C (\omega_{\text{fb}} , T_{{1}}) + f_{3}(T_{{2}})  C (\omega_{\text{fb}} , T_{{2}}) \right], 
	\label{eq:variance}
\end{equation}
with the function $f_{x}(T) = 1 + x(T/T_{\text{h}})^{2}$ and the heat capacity, $C (\omega, T) = \partial_{T} \text{tr} [ \rho (\omega, T) H  ]$,  of the working substance in a thermal state. The above analytical expressions are exact in the low dissipation regime. We shall now use them to investigate the power-efficiency-stability trade-off of the quantum information machine.

\paragraph{Power-efficiency-stability  trade-off.}
Two quantities are usually introduced to quantify the relative fluctuations of a random observable: the Fano factor, $\sigma_{w}^{2}/{|W|}$, defined as the ratio of variance to mean, and the coefficient of variation, $\sigma_{w}/{|W|}$, defined as the quotient of standard deviation and mean \cite{eve98}. The advantage of the Fano factor is that it is equal to one for Poissonian stochastic variables, while the coefficient of variation is dimensionless. We shall employ both in the following.

Starting with Eq.~\eqref{eq:variance},  the product of the  Fano factor and the efficiency,  $\eta = 1 - \Sigma_{\text{h}} / \Delta S \tau_{\text{h}}$, can be written as
\begin{equation}
	\frac{\sigma_{w}^{2}}{|W|} \eta = \frac{T_{\text{h}} \left[ f_{5} (T_{{1}})  C (\omega_{\text{fb}} , T_{{1}}) + f_{3}(T_{{2}})  C (\omega_{\text{fb}} , T_{{2}}) \right]}{ \Delta S }.
	\label{eq:fanoEff}
\end{equation}
We note that   the right-hand side  is a constant that does not depend on the cycle time. Equation \eqref{eq:fanoEff} hence shows that the Fano factor is inversely proportional to the efficiency. The relative work fluctuations  are consequently finite, especially in the limit  $\tau_{\text{h}}\to \infty$, where the efficiency $\eta$ achieves its maximal value of one, which corresponds to  perfect information-to-work conversion  \cite{qua06,jac09,kim11,str13,bra15,elo17,elo18,sea20}. However, in that limit the power,  $\mathcal{P} = |W| / (\tau_{\text{h}} + \tau_{\text{fb}})$, vanishes.  We shall therefore focus on the situation where there is a clear timescale separation between the internal relaxation time of the working substance and the (much shorter) thermalization time induced by the coupling strength to the heat reservoir. In order to examine the performance of the machine in the limit of  fast thermalization time, we  introduce a dimensionless scaling parameter $\kappa \in \mathbb{R}_{>0}$, and rescale expansion time, $\tau_{\text{h}} \to \tau_{\text{h}} /\kappa^{\alpha}$, and dissipation, $\Sigma_{\text{h}} \to \Sigma_{\text{h}} / \kappa^{\gamma}$, with 
$\gamma > \alpha >0 $, such that $(\Sigma_{\text{h}} / \kappa^{\gamma}) / (\tau_{\text{h}} /\kappa^{\alpha}) \to 0$ when $\kappa \to \infty$. The rescaling of the dissipation $\Sigma_{\text{h}}$ is achieved by rescaling  the coupling strength between the working system and the heat bath, so that the relaxation time of the system is always shorter than $\tau_{\text{h}} / \kappa^{\alpha}$ (an explicit example of the relation between $\Sigma_{\text{h}}$ and the coupling  for a two-level system may be found in  Ref.~\cite{fal23}). In the limit $\kappa \to \infty$, we  concretely obtain for efficiency,  power and Fano factor of the quantum information engine (Supplemental Material)
\begin{eqnarray}
\label{4}
		&& \lim_{\kappa \to \infty} \eta = 1, \qquad 
		 \lim_{\kappa \to \infty} \mathcal{P} = \frac{T_{\text{h}} \Delta S}{\tau_{\text{fb}}} \qquad \text{ and }\qquad  \\
		&&\lim_{\kappa \to \infty} \frac{\sigma_{w}^{2}}{|W|} = \frac{T_{\text{h}} \left[ f_{5} (T_{{1}})  C (\omega_{\text{fb}} , T_{{1}}) + f_{3}(T_{{2}})  C (\omega_{\text{fb}} , T_{{1}}) \right]}{ \Delta S }. \nonumber	
\end{eqnarray}
Equations \eqref{4} indicate that the limit of  fast thermalization leads to a Carnot information engine with maximum efficiency, finite power output and (small) work output fluctuations.  In this regime, the quantum machine hence operates as a  stable engine \cite{pie18}. Figure \ref{fig:scaling} displays the behavior of the above quantities, including the mean work and the work variance \eqref{eq:variance}, as well as the efficiency-fluctuation trade-off equality \eqref{eq:fanoEff}, for a two-level system with $\mathcal{P} = \text{diag}(2, 1)$, as a function of the scaling exponent $\kappa$. We observe that both efficiency and power increase with $\kappa$, whereas the Fano factor steadily decreases. Interestingly, efficiency and Fano factor reach their respective asymptotic values (dotted lines)  very quickly, even for moderate values of $\kappa$, following a power law $\kappa^{ \alpha-\gamma}$; the same holds true for mean and variance. The power further approaches its asymptotic limit with either a smaller exponent as  $\kappa^{ - \alpha}$ if $\gamma \geq 2 \alpha$ (as in Fig.~\ref{fig:scaling}), or as $\kappa^{ \alpha-\gamma}$ otherwise. This means that the mathematical limit of $\kappa \to \infty$ is not strictly necessary, which is a crucial point for experimental investigations of such devices.

\begin{figure*}[t]
    \begin{center}
    \includegraphics[scale=.52]{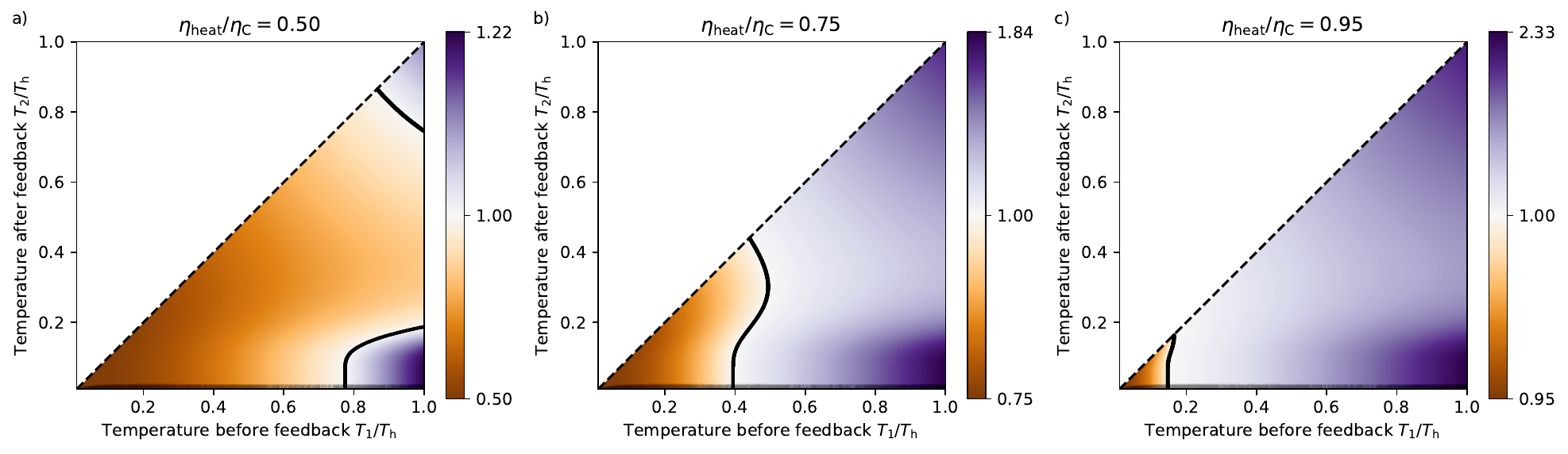}
    \caption{Relative work fluctuations of finite-time Carnot information and heat engines. Quotient of the respective coefficients of variations, $\text{COV}_{\text{info}} / \text{COV}_{\text{heat}}$, Eq.~(7),  a function of the temperatures before and after the measurement and feedback protocol, $T_{1}$ and $T_{2}$, for a two-level system, and for various values of the efficiency $\eta_{\text{heat}} / \eta_{\text{C}} $. The information engine is more stable than the heat engine in the orange shaded areas, which decrease  as the efficiency approaches the Cartnot efficiency $\eta_\text{C}$. The solid black line indicates the values of the temperatures such that $\text{COV}_{\text{info}} / \text{COV}_{\text{heat}} = 1$, while the shaded region at the bottom corresponds to the values that saturate the upper bound (7). Parameters  are $\omega_{\text{fb}} = 1$ and $\kappa = 1$.}
    \label{fig:quotient}
\end{center}
\end{figure*}

\paragraph{Heat versus information engine.}
We  proceed by comparing the relative work fluctuations of  the finite-time quantum Carnot information engine and those of the standard quantum Carnot heat engine \cite{qua07,abe11,dan20,abi20,den21}. We focus on the coefficient of variation, $\text{COV} = \sigma_{w} /  |W| $, because it provides simpler expressions (and set $\kappa = 1$). In order to make the comparison  as fair as possible, the only difference between the two must be the measurement-and-feedback branch which replaces the cold isothermal compression. We  therefore assume that the hot reservoir and the state of the working substance before and after the hot isothermal expansion are  the same in both cases. This has two implications: (i) first, the heat  $Q_{\text{h}}$ and the work  $W_\text{h}$ produced during the hot isothermal compression are identical in both cycles, and (ii)  second, the frequencies $\omega_{1}$ and $\omega_{2}$, of the working medium  before and after the cold isothermal compression at temperature $T_{\text{c}}$ satisfy $\omega_{1} / T_{\text{c}} = \omega_{\text{fb}} / T_{{1}}$ and $\omega_{2} / T_{\text{c}} = \omega_{\text{fb}} / T_{{2}}$   (since the  branches (2-3) and (4-1) are isentropic). Under these conditions, the coefficient of variation of the finite-time quantum Carnot heat engine is given by  \cite{den21}
\begin{equation}
	\text{COV}_{\text{heat}} = \eta_{\text{C}} T_{\text{h}} \frac{ \sqrt{ C (\omega_{\text{fb}} , T_{{1}}) + C (\omega_{\text{fb}} , T_{{2}}) } }{Q_{\text{c}} + Q_{\text{h}}},
	\label{eq:COVT}
\end{equation}
where $Q_{\text{c}} = T_{\text{c}} (- \Delta S - \Sigma_{\text{c}} / \tau_{\text{c}})$ is the heat exchanged during the cold isothermal compression, and $\Sigma_{\text{c}}$ characterizes the entropy production during time  $\tau_{\text{c}}$. On the other hand, following Eq.~\eqref{eq:variance}, the coefficient of variation of the finite-time quantum Carnot information cycle reads
\begin{equation}
	\text{COV}_{\text{info}} = T_{\text{h}} \frac{ \sqrt{ f_{5} (T_{{1}})  C (\omega_{\text{fb}} , T_{{1}}) + f_{3} (T_{{2}})  C (\omega_{\text{fb}} , T_{{2}}) } }{Q_{\text{h}}}.
	\label{eq:COVI}
\end{equation}
 By noticing that $ f_{3} (T_{{2}}) \leq f_{5} (T_{{1}})$, we can derive universal lower and upper bounds for the quotient  of the two coefficients of variation, Eqs.~\eqref{eq:COVT} and \eqref{eq:COVI}, that do not depend on the working substance performing the cycle:
 \begin{equation}
	\sqrt{f_{3} (T_{{2}})}  \frac{\eta_{\text{heat}}}{\eta_{\text{C}}} \leq  \frac{\text{COV}_{\text{info}}}{\text{COV}_{\text{heat}}} \leq \sqrt{f_{5} (T_{{1}})}  \frac{\eta_{\text{heat}}}{\eta_{\text{C}}},
	\label{eq:COVbounds}
\end{equation}
where $\eta_{\text{heat}} = 1 + Q_{\text{c}} / Q_{\text{h}}$ is the efficiency of the finite-time Carnot heat engine  \cite{esp10}. In general, $\eta_{\text{heat}} \leq \eta_{\text{C}}$ and $\eta_{\text{heat}} \to \eta_{\text{C}}$ in the infinite-time (reversible) limit  $(\tau_{\text{c}} , \tau_{\text{h}}) \to \infty$. Notably, the lower bound in Eq.~\eqref{eq:COVbounds} is smaller than $1$ if $\eta_{\text{heat}} \leq \eta_{\text{C}} / \sqrt{f_{3} (T_{{2}})}$, indicating the possibility of a finite-time regime in which the relative work fluctuations of the  information engine are smaller than those of the corresponding heat engine (the lower bound is always larger than one in the infinite-time limit $\eta_{\text{heat}} \to \eta_{\text{C}}$). This implies a more stable information machine with larger work output and/or smaller work output fluctuations. In addition, the existence of an upper bound means that the relative work fluctuations of the Carnot information cycle cannot be arbitrarily larger than those of the corresponding heat engine. We further note that the upper bound is tight, when the measurement-and-feedback scheme cools the system to the ground state: $\text{COV}_{\text{info}} / \text{COV}_{\text{heat}} \to \sqrt{f_{5} (T_{{1}})} \eta_{\text{heat}} / \eta_{\text{C}}$, when $T_{{2}} \to 0$ and $C (\omega_{\text{fb}} , T_{{2}}) \to 0$. Since the measurement-and-feedback control in the Carnot information engine introduces extra work fluctuations compared to the cold isothermal compression in the thermal cycle \cite{den21}, the decrease of  the relative work fluctuations  can be explained by an increase of the average work output. Indeed, since an information reservoir only exchanges information with the system, but no energy \cite{def13,bar14,bar14a}, the net energy balance of the measurement-and-feedback branch is zero, whereas it is negative  in the cold isothermal expansion of the heat engine. This underscores a fundamental difference, and a  fundamental thermodynamic advantage, of the information cycle compared to its thermal counterpart.

Figure~\ref{fig:quotient} presents the ratio $\text{COV}_{\text{info}} / \text{COV}_{\text{heat}}$ for a two-level working medium, as a function of the two temperatures $T_{{1}}$ and $T_{{2}}$, before and after the measurement-and-feedback step, for different values of $\eta_{\text{heat}} / \eta_{\text{C}} $. {The upper bound shown in Eq.~\eqref{eq:COVbounds} is always reached.} For $\eta_{\text{heat}} / \eta_{\text{C}} =0.5$, the information engine is more stable than the heat engine for most of the parameter space (orange areas). These regions shrink when $\eta_{\text{heat}}$ approaches the (infinite-time) Carnot efficiency $\eta_{\text{C}}$. However, remarkably, even when the quantum Carnot  heat engine operates almost reversibly, $\eta_{\text{heat}} / \eta_{\text{C}} = 0.95$, there is a set of parameters for which the relative work fluctuations of the quantum Carnot information engine are smaller.

\paragraph{Conclusions.} We have  analyzed the power-efficiency-stability trade-off in a finite-time quantum Carnot information engine which converts information into work. We have found that the Fano factor, which quantifies relative work output fluctuations, is inversely proportional to the efficiency. Moreover, maximum efficiency can be reached at nonzero power and finite (even small) work output fluctuations in the limit of fast thermalization. Interestingly, the finite-time Carnot information engine can be more stable than the associated Carnot heat engine, with smaller relative work output fluctuations. Our findings show that producing work via measurement and feedback (in the presence of one heat bath), in a finite-time cyclic quantum Maxwell demon setting, can be more advantageous thermodynamically than converting heat into work (in the presence of two heat baths). Such high-efficiency, high-power, high-stability quantum machines thus offer a practical alternative to conventional heat engines. \\

\textit{Acknowledgements}. Financial support from the German Research Foundation (DFG) (FOR 2427) is acknowledged. The authors further thank Franklin L. S. Rodrigues, Cyril Elouard and Karen V. Hovhannisyan for helpful discussions and advice.

\clearpage
\widetext
\begin{center}
\textbf{\large Supplemental Material:  Power-efficiency-stability trade-off in quantum information engines}
\end{center}
\setcounter{equation}{0}
\setcounter{figure}{0}
\setcounter{table}{0}
\setcounter{page}{1}
\setcounter{secnumdepth}{4}
\makeatletter
\renewcommand{\theequation}{S\arabic{equation}}
\renewcommand{\thefigure}{S\arabic{figure}}
\renewcommand{\bibnumfmt}[1]{[S#1]}
\renewcommand{\citenumfont}[1]{S#1}


\renewcommand{\figurename}{Supplementary Figure}
\renewcommand{\theequation}{S\arabic{equation}}
\renewcommand{\thefigure}{S\arabic{figure}}
\renewcommand{\bibnumfmt}[1]{[S#1]}
\renewcommand{\citenumfont}[1]{S#1}

\section{Derivation of the work output probability distribution}

In this section, we will derive the work probability distribution $P(w)$ of the quantum Carnot information engine, given in Eq.~(1) in the main text, using the end-point measurement scheme \cite{ghe21,gia23,her23}. 

\subsection{End-point measurement scheme}

Three main measurement methods are commonly employed to evaluate the statistics of the energy change during a quantum process: 1) the end-point measurement scheme (EPM) \cite{ghe21,gia23,her23}, the two-point measurement approach (TPM)  \cite{tal07,esp09a,cam11} and dynamic Bayesian networks (DBN) \cite{nea03,dar09,mic20,par20,mic24}. We shall here compare the three techniques, and show that only the first one does not affect the known energy balance of the measurement-and-feedback step of the finite-time quantum Carnot information cycle \cite{fad23}. 

Let us consider an arbitrary CPTP map $\Phi$ that takes some initial state $\rho_{\text{i}}$ to a final state $\rho_{\text{f}} = \Phi (\rho_{\text{i}})$, where the Hamiltonian of the system  goes from $H_{\text{i}}$ to $H_{\text{f}}$. We are interested in the stochastic energy change during the process, so we want to compute the corresponding probability distribution energy difference
\begin{equation}
	P(\Delta u) = \sum_{j,k} \delta \big[ \Delta u - ( e_{\text{f}}^{k} - e_{\text{i}}^{j} ) \big] P (e_{\text{i}}^{j}, e_{\text{f}}^{k}),
\end{equation}
where $e_{\text{i (f)}}^{j (k)}$ is the $j$($k$)-th eigenvalue of the initial (final) Hamiltonian $H_{\text{i (f)}}$, and $P (e_{\text{i}}^{j}, e_{\text{f}}^{k})$ is the joint probability distribution of obtaining such eigenvalues. The expression of $P (e_{\text{i}}^{j}, e_{\text{f}}^{k})$, depends on the chosen measurement scheme.

The most commonly used scheme to compute $P (e_{\text{i}}^{j}, e_{\text{f}}^{k})$ is TPM \cite{tal07,esp09a,cam11}. As its name indicates, this method consists in performing two projective energy measurements: one at the beginning and the other one at the end of the process. When the first measurement is performed, the initial state $\rho_{\text{i}}$ collapses to $\Pi_{\text{i}}^{j} / \text{tr} (\Pi_{\text{i}}^{j})$, where $\Pi_{\text{i}}^{j}$ is the projector on the eigenspace corresponding to the eigenvalue $e_{\text{i}}^{j}$ of $H_{\text{i}}$. Then, the post-measurement collapsed state evolves and a final projective measurement is performed on $\Phi [\Pi_{\text{i}}^{j} / \text{tr} (\Pi_{\text{i}}^{j})]$. The resulting distribution is
\begin{equation}
	P_{\text{TPM}} (e_{\text{i}}^{j}, e_{\text{f}}^{k}) = P (e_{\text{i}}^{j}) P ( e_{\text{f}}^{k} | e_{\text{i}}^{j}) = \text{tr} ( \Pi_{\text{i}}^{j} \rho_{\text{i}} ) \, \text{tr} \{ \Pi_{\text{f}}^{k} \Phi [\Pi_{\text{i}}^{j} / \text{tr} (\Pi_{\text{i}}^{j})] \}.
\end{equation}
An obvious drawback of the scheme is that the initial measurement destroys all the coherences of the initial state in the energy basis. As a consequence, the expectation value of $\Delta u$ in general does not coincide with $\Delta E = \text{tr} (\rho_{\text{f}} H_{\text{f}}) - \text{tr} (\rho_{\text{i}} H_{\text{i}})$ in the presence of coherence. Since the states before and after the measurement-and-feedback step are (incoherent) thermal states, this issue does not affect the present study. Furthermore, the actual statistics of $\rho_{\text{f}}$ are lost because the final state considered is no longer $\Phi (\rho_{\text{i}})$ but rather an evolved projector $\Phi [\Pi_{\text{i}}^{j} / \text{tr} (\Pi_{\text{i}}^{j})]$.

The dynamical Bayesian network scheme improves on some of the drawbacks of the TPM scheme at the cost of knowing more information about the initial state and the evolution \cite{nea03,dar09,mic20,par20,mic24}. While the TPM scheme is state- and process-independent, meaning that the implementation is always the same regardless of $\rho_{\text{i}}$ and $\Phi$, the DBN scheme is not. In this case, we need to know the eigenbasis of $\rho_{\text{i}}$ and how $\Phi$ evolves each of the corresponding projectors. Consider starting the measurement process with not one, but rather two preparations of the initial state: $\rho_{\text{i}} \otimes \rho_{\text{i}}$. First, each preparation is measured in the state eigenbasis with the operator $\Pi_{\text{i}}^{s} \otimes \Pi_{\text{i}}^{s}$. Then, the first preparation is measured in the eigenbasis of $H_{\text{i}}$, and the second one is left to evolve and then measured in the eigenbasis of $H_{\text{f}}$. The joint probability distribution in this case is given by the sum over all possible trajectories:
\begin{equation}
	P_{\text{DBN}} (e_{\text{i}}^{j}, e_{\text{f}}^{k}) = \sum_{s} P_{s} P(e_{\text{i}}^{j} | s) P(e_{\text{f}}^{k} | s) = \sum_{s} \text{tr} (\Pi_{\text{i}}^{s} \rho_{\text{i}}) \, \text{tr} \{ \Pi_{\text{i}}^{j} [ \Pi_{\text{i}}^{s} / \text{tr} (\Pi_{\text{i}}^{s} )] \} \, \text{tr} \{ \Pi_{\text{f}}^{k} \Phi [ \Pi_{\text{i}}^{s} / \text{tr} (\Pi_{\text{i}}^{s}) ] \}.
\end{equation}
It is straightforward to check that, contrary to TPM, the expectation value of $\Delta u$ always coincides with $\Delta E$, so the DBN framework preserves the coherences of the initial state. If the initial state was already diagonal in the energy eigenbasis, then $P_{\text{DBN}} = P_{\text{TPM}}$. Nevertheless, if suffers from the same drawback of considering as a final state an evolved projector $\Phi [ \Pi_{\text{s,i}} / \text{tr} (\Pi_{s,\text{i}}) ]$ instead of $\rho_{\text{f}}$.

In both methods, TPM and DBN, we saw that the final state considered in the measurement scheme is not actually $\rho_{\text{f}}$ but rather an evolved projector, consequence of the initial measurement on $\rho_{\text{i}}$.  This is specially problematic for the analysis of the energy fluctuations in the Carnot information engine due to the presence of the state-dependent measurement-and-feedback operation. Indeed, let us consider any of those schemes applied to the measurement-and-feedback step. The initial state could be any of the $\rho_{i} = M_{i} \rho (\omega_{\text{fb}}, T_{\text{b}}) M_{i}^{\dagger} / \text{tr} [M_{i} \rho (\omega_{\text{fb}}, T_{{1}}) M_{i}^{\dagger}]$ resulting from the previous measurement step in the cycle. For each of those states, the feedback controller applies a map $\Phi_{i}$ that acts as $\Phi_{i} (\rho_{i}) = \rho (\omega_{\text{fb}}, T_{{2}})$, so the final state is always the same. Now, suppose that the first measurement that the scheme requires is performed on the initial state $\rho_{i}$ and it collapses to some projector $\Pi_{j}$. Since the map $\Phi_{i}$ that the feedback controller employs depends on the state that it will be applied to (and on the result of the measurement step in the cycle), and the controller does not have any map for $\Pi_{j}$, the process halts and the cycle cannot continue. One could potentially get around the problem by arguing that the feedback controller is smart enough to know what to do, and given \textit{any} input state it follows the general recipe of reordering the eigenvalues and shifting the energy levels with unitary operations so that the final state is always $\rho (\omega_{\text{fb}}, T_{{2}})$. But, if the state that the controller receives is a projector, then it cannot transform it into $\rho (\omega_{\text{fb}}, T_{{2}})$ with unitary operations unless $T_{{2}} = 0$. That is, the outcome of the feedback process after collapsing the input state will inevitably be the ground state: $\Phi (\Pi_{j}) = \lvert 0 \rangle \langle 0 \rvert$ for any projector $\Pi_{j}$. Therefore, the expectation value of $\Delta u$ will be $\langle \Delta u \rangle = \langle 0 \rvert H_{\text{f}} \lvert 0 \rangle - \text{tr} (\rho_{i} H_{\text{i}})$, which is \textit{always} different from the actual energy change $\Delta E = \text{tr} [\rho (\omega_{\text{fb}}, T_{{2}}) H_{\text{f}}] - \text{tr} (\rho_{i} H_{\text{i}})$ (unless $T_{{2}} = 0$). We may hence conclude that both TPM and DBN negatively affect the measurement-and-feedback step such that the measured information cycle is no longer a quantum Carnot information engine. The two methods are therefore not suitable to properly analyze the energy fluctuations of the Carnot information cycle. It is worth noting that this is not unique to TPM and DBN; every scheme that collapses the state that is going to evolve afterwards also presents the same problem.

A measurement technique that does not suffer from the above issues is the EPM scheme \cite{ghe21,gia23,her23}, which has been experimentally implemented in Ref.~\cite{her23}. Operationally, it can be thought of as similar to DBN but without the initial collapse in the eigenbasis of $\rho_{\text{i}}$. That is, $\rho_{\text{i}} \otimes \rho_{\text{i}}$ is initially prepared, and a measurement is performed in the energy basis on one of the states. Then, the other one is left to evolve unperturbed, and a second measurement is performed on the final state. The probability distribution is thus constructed as
\begin{equation}
	P_{\text{EPM}} (e_{\text{i}}^{j}, e_{\text{f}}^{k}) = P (e_{\text{i}}^{j}) P ( e_{\text{f}}^{k}) = \text{tr} ( \Pi_{\text{i}}^{j} \rho_{\text{i}} ) \, \text{tr} [\Pi_{\text{f}}^{k} \Phi (\rho_{\text{i}}) ].
\end{equation}
This scheme presents two advantages compared to TPM and DBN that are essential for the  analysis of the energy fluctuations of the information engine cycle. First, the expectation value of $\Delta u$ \textit{always} matches the energy difference $\Delta E$: $\langle \Delta u \rangle = \Delta E$ for any initial and final states. And second, it does not disturb the evolution, so the feedback controller can do its job properly, implying that  the statistics of the genuine final state $\rho_{\text{f}} = \Phi (\rho_{\text{i}})$ are recovered. 

\subsection{Computation of the distribution $P(w)$}

Before starting with the computation of the work probability distribution $P (w)$ using EPM, we should note that, due to the very nature of this scheme, the probability distributions involved are all independent of each other. There are no conditional probability distributions. Therefore, we will analyze each branch of the cycle separately. Furthermore, we will split the measurement step from the feedback step, and analyze each on its own as to properly account for the fluctuations introduced by them.

As done in the main text, the total stochastic work is decomposed in the work produced by the working medium plus the work extracted by the measurement-and-feedback protocol, $w = w_{\text{wm}} + w_{\text{fb}}$, and the corresponding probability distributions $P_{\text{wm}}$ and $P_{\text{fb}}$, respectively, are computed. The total distribution is thus given by the convolution $P = P_{\text{wm}} \ast P_{\text{fb}}$. 

Let us begin with $P_{\text{wm}}$. The work produced by the working medium is the work corresponding to the isentropic compression, isothermal expansion, and isentropic expansion branches of the engine cycle. It thus can be written as $w_{\text{wm}} = w_{23} + w_{34} + w_{41}$, where $w_{ij}$ denotes the stochastic work extracted when the working medium goes from step $i$ to step $j$ (see Fig. 1 in the main text) and each corresponding $P_{ij}$ can be computed. The total probability distribution is then $P_{\text{wm}} = P_{23} \ast P_{34} \ast P_{41}$. The probability distributions for the work extracted during the isentropic processes, $P_{23}$ and $P_{41}$, are just the probability distributions of (minus) the internal energy change, since the working medium is not in contact with any heat bath. Using the EPM framework described above, they are 
\begin{equation}
	P_{23} (w_{23}) = \sum_{j,k} \delta \big[ w_{23} + (E_{3}^{k} - E_{2}^{j}) \big] p_{3}^{k} p_{2}^{j}
\end{equation}
and
\begin{equation}
	P_{41} (w_{41}) = \sum_{j,k} \delta \big[ w_{41} + (E_{1}^{k} - E_{4}^{j}) \big] p_{1}^{k} p_{4}^{j},
\end{equation}
where $p_{i}^{j} = \text{tr} [ \Pi_{i}^{j}  \, \rho(\omega_{i} , T_{i})]$ is the probability of obtaining the eigenvalue $E_{i}^{j}$ of $H_{i}$, with $\Pi_{i}^{j}$ the projector over the corresponding eigenspace. As for the probability distribution for the work extracted during the isothermal expansion, $P_{34}$, in the limit in which the process is done quasistatically, work is deterministic and, therefore, the probability distribution is sharp  \cite{dan20,den21}:
\begin{equation}
 	P_{34} (w_{34}) = \delta ( w_{34} - W_{\text{h}}).
\end{equation}
The work output according to the first law of thermodynamics is $W_{\text{h}} = Q_{\text{h}} - \Delta U_{\text{h}}$, where the change of the internal energy of the working substance is given by $\Delta U_{\text{h}} = \text{tr} [ \rho (\omega_{4} , T_{\text{h}}) H_{4} ] - \text{tr} [ \rho (\omega_{3} , T_{\text{h}}) H_{3} ]$. $P_{\text{wm}}$ can now be written as
\begin{equation}
	P_{\text{wm}} (w_{\text{wm}}) = \sum_{j,k} \sum_{l,m} \delta \big[ w_{\text{wm}} + (E_{3}^{k} - E_{2}^{j}) - W_{\text{h}} +  (E_{1}^{m} - E_{4}^{l}) \big] p_{3}^{k} p_{2}^{j} p_{1}^{m} p_{4}^{l}.
	\label{eqsup:pWM}
\end{equation}
The mean value $\langle w_{\text{wm}} \rangle$ is easily computed, and the result is
\begin{equation}
	\langle w_{\text{wm}} \rangle = \int dw \, w \, P_{\text{wm}} (w)  = - \Delta U_{23} + W_{\text{h}} - \Delta U_{41} = Q_{\text{h}} + \Delta U_{12}.
\end{equation}
Viewed from a thermodynamic perspective, without making reference to the measurement and corresponding feedback, the process $1 \to 2$ is an isochoric transformation during which  no work is performed. Thus, the energy difference $\Delta U_{12}$ is equal to the  heat exchanged with the information reservoir $Q_{\text{c}}$, and we obtain $\langle w_{\text{wm}} \rangle = Q_{\text{h}} + Q_{\text{c}}$ \cite{fad23}.

We now proceed to the computation of the probability distribution corresponding to the work extracted from the measurement-and-feedback operation, $P_{\text{fb}}$. From the point of view of the external controller, it invests a stochastic amount of energy $e_{\text{m}}$ to perform the measurement and another amount $e_{\text{f}}$ to implement the feedback operation. Since the process is specifically chosen to be thermodynamically reversible \cite{fad23}, the work extracted is simply equal to (minus) the energy invested. Consequently, we can write $w_{\text{fb}} = - (e_{\text{m}} + e_{\text{f}})$, with probability distribution $P_{\text{fb}} = \tilde{P}_{\text{m}} \ast \tilde{P}_{\text{f}}$, where $\tilde{P} (w) = P (-w)$. After averaging over all possible measurement outcomes $\rho_{i}$ and using the fact that $\sum_{i} M_{i}^{\dagger}  M_{i}= \mathbb{1}$, the probability distributions are
\begin{equation}
	P_{\text{m}} ( e_{\text{m}} ) = \sum_{j, k} \delta \big[ e_{\text{m}} - (E_{1}^{k} - E_{1}^{j}) \big] p_{1}^{k}  p_{1}^{j},
\end{equation}
and
\begin{equation}
	P_{\text{f}} ( e_{\text{f}} ) = \sum_{j, k} \delta \big[ e_{\text{f}} - (E_{2}^{k} - E_{1}^{j}) \big] p_{2}^{k}  p_{1}^{j}.
\end{equation}
Therefore, the complete probability distribution from the controller side is
\begin{equation}
	P_{\text{fb}} (w_{\text{fb}}) = \sum_{j,k} \sum_{l,m} \delta \big[w_{\text{mf}} + (E_{1}^{k} - E_{1}^{j}) + (E_{2}^{m} - E_{1}^{l}) \big] p_{1}^{k}  p_{1}^{j} p_{2}^{m}  p_{1}^{l}.
	\label{eqsup:pMF}
\end{equation}
Notably, $P_{\text{fb}}$ is independent of the particular choice of measurement operators $\{ M_{i} \}$ as long as they are diagonal in the energy basis. Furthermore, the measurement-and-feedback process introduces fluctuations in the cycle. This is in stark contrast to an isothermal compression in which work is deterministic if done adiabatically. The average $\langle w_{\text{fb}} \rangle$ is
\begin{equation}
	\langle w_{\text{fb}} \rangle = \int dw \, w \,  P_{\text{fb}} (w) = - \Delta U_{12},
\end{equation}
which coincides with the one obtained in Ref.~\cite{fad23}.

Finally, combining Eqs. \eqref{eqsup:pWM} and \eqref{eqsup:pMF}, the full probability distribution is obtained:
\begin{equation}
	P (w) = \sum_{j,k,l,m} \sum_{p,q,r,s}  \delta \big[ w + (E_{3}^{k} - E_{2}^{j}) - W_{\text{h}} +  (E_{1}^{m} - E_{4}^{l}) + (E_{1}^{q} - E_{1}^{p}) + (E_{2}^{s} - E_{1}^{r}) \big] p_{3}^{k} p_{2}^{j} p_{1}^{m} p_{4}^{l} p_{1}^{q}  p_{1}^{p} p_{2}^{s}  p_{1}^{r}.
\end{equation}
The mean value of the total work extracted during the cycle is accordingly
\begin{equation}
	\langle w \rangle = \int dw \, w \, P (w) = \langle w_{\text{wm}} \rangle + \langle w_{\text{fb}} \rangle = Q_{\text{h}}.
\end{equation}

\section{Work output fluctuations}

In this section, we are going to compute the variance of the work output, $\sigma_{w}^{2} = \langle w^{2} \rangle - \langle w \rangle^{2}$. Since the probability distributions for each branch are independent of each other, the total variance is just the sum of the individual variances. Therefore,
\begin{equation}
	\begin{aligned}
		\sigma_{w}^{2}  = \sigma_{\text{wm}}^{2} + \sigma_{\text{fb}}^{2}  
				       = \sigma_{23}^{2} + \sigma_{34}^{2} + \sigma_{41}^{2} + \sigma_{\text{m}}^{2} + \sigma_{\text{f}}^{2} 
				       = 4 \sigma_{1}^{2} + 2 \sigma_{2}^{2} + \sigma_{3}^{2} + \sigma_{4}^{2},
	\end{aligned}
\end{equation}
where $\sigma_{i}^{2} = \text{tr} [\rho(\omega_{i} , T_{i}) H_{i}^{2}] - \text{tr} [\rho(\omega_{i} , T_{i}) H_{i}]^{2}$ is the variance of the energy in the state $\rho(\omega_{i} , T_{i})$. Using the fact that  the heat capacity for a thermal state is related to the energy fluctuations as $C (\omega_{i}, T_{i}) = \sigma_{i}^{2} / T_{i}^{2}$ \cite{blu06}, we can rewrite the above expression as
\begin{equation}
	\sigma_{w}^{2} = 4 T_{1}^{2} C (\omega_{1}, T_{1}) + 2 T_{2}^{2} C (\omega_{2}, T_{2}) + T_{\text{h}}^{2} C (\omega_{3}, T_{\text{h}}) +  + T_{\text{h}}^{2} C (\omega_{4}, T_{\text{h}}).
\end{equation}
Since the states are thermal with a Hamiltonian $H_{\text{t}} = \omega_{\text{t}} \mathcal{P}$, the heat capacity is a function of the quotient $\omega / T$, and not of $\omega$ and $T$ separately. Therefore, if $\omega_{i} / T_{i} = \omega_{j} / T_{j}$, $C (\omega_{i}, T_{i}) = C (\omega_{j}, T_{j})$. Because $\rho (\omega_{2} , T_{2})$ and $\rho (\omega_{3} , T_{\text{h}})$, and $\rho (\omega_{4} , T_{\text{h}})$ and $\rho (\omega_{1} , T_{1})$ are related via unitary transformations, we have $\omega_{2} / T_{2} = \omega_{3} / T_{\text{h}}$ and $\omega_{1} / T_{1} = \omega_{4} / T_{\text{h}}$.  We eventually arrive at
\begin{equation}
	\sigma_{w}^{2} = T_{\text{h}}^{2} \left[ f_{5} (T_{1}) C (\omega_{1}, T_{1}) + f_{3} (T_{2}) C (\omega_{2}, T_{2})  \right],
\end{equation}
where $f_{x} (T) = 1 + x (T / T_{\text{h}})^{2}$. 

\section{Asymptotic expressions for  efficiency, power and  Fano factor}

Employing the scaling detailed in the main text, where the expansion time in the hot isotherm is replaced by $\tau_{\text{h}} \to \tau_{\text{h}} / \kappa^{\alpha}$ and the coupling is changed so that the dissipation changes as $\Sigma_{\text{h}} \to \Sigma_{\text{h}} / \kappa^{\gamma}$, with $\kappa , \alpha, \gamma \in \mathbb{R}_{>0}$ and $\gamma > \alpha$, we obtain  the following efficiency, power and Fano factor: 
\begin{equation}
	\eta = 1 - \frac{\Sigma_{\text{h}}}{\Delta S \, \tau_{\text{h}}} \quad \longrightarrow \quad \eta = 1 - \frac{1}{\kappa^{\gamma - \alpha}} \frac{\Sigma_{\text{h}}}{\Delta S \, \tau_{\text{h}}},
\end{equation}
\begin{equation}
	\mathcal{P} = \frac{T_{\text{h}}}{\tau_{\text{fb}} + \tau_{\text{h}}} \left( \Delta S - \frac{\Sigma_{\text{h}}}{\tau_{\text{h}}} \right) \quad \longrightarrow \quad \mathcal{P} = \frac{T_{\text{h}}}{\tau_{\text{fb}} + \tau_{\text{h}} / \kappa^{\alpha}} \left( \Delta S - \frac{1}{\kappa^{\gamma - \alpha}} \frac{\Sigma_{\text{h}}}{\tau_{\text{h}}} \right),
\end{equation}
and
\begin{equation}
	\frac{\sigma_{w}^{2}}{|W|} = \frac{T_{\text{h}} \left[ f_{5} (T_{\text{b}})  C (\omega_{\text{fb}} , T_{\text{b}}) + f_{3}(T_{\text{a}})  C (\omega_{\text{fb}} , T_{\text{a}}) \right]}{ \Delta S - \frac{\Sigma_{\text{h}}}{\tau_{\text{h}}} } \quad \longrightarrow \quad \frac{\sigma_{w}^{2}}{|W|} = \frac{T_{\text{h}} \left[ f_{5} (T_{\text{b}})  C (\omega_{\text{fb}} , T_{\text{b}}) + f_{3}(T_{\text{a}})  C (\omega_{\text{fb}} , T_{\text{a}}) \right]}{ \Delta S - \frac{1}{\kappa^{\gamma - \alpha}}  \frac{\Sigma_{\text{h}}}{\tau_{\text{h}}} }.
\end{equation}
If we perform a large $\kappa$ expansion in the power and the Fano factor, we further find
\begin{equation}
	\mathcal{P} \simeq \frac{T_{\text{h}} \Delta S}{\tau_{\text{fb}}} \left( 1 - \frac{1}{\kappa^{\alpha}} \frac{\tau_{\text{h}}}{\tau_{\text{fb}}} - \frac{1}{\kappa^{\gamma - \alpha}}  \frac{\Sigma_{\text{h}}}{\Delta S \, \tau_{\text{h}}} \right),
\end{equation}
and
\begin{equation}
	\frac{\sigma_{w}^{2}}{|W|} \simeq \frac{T_{\text{h}} \left[ f_{5} (T_{{1}})  C (\omega_{\text{fb}} , T_{{1}}) + f_{3}(T_{{2}})  C (\omega_{\text{fb}} , T_{{2}}) \right]}{ \Delta S } \left( 1 +  \frac{1}{\kappa^{\gamma - \alpha}}  \frac{\Sigma_{\text{h}}}{\Delta S \, \tau_{\text{h}}} \right).
\end{equation}
Therefore, for large $\kappa$, both the efficiency and the Fano factor approach their limiting values as $1 / \kappa^{\gamma - \alpha}$, but the power does it as $1 / \kappa^{\gamma - \alpha}$ if $\gamma < 2 \alpha$, or as $1 / \kappa^{\alpha}$ otherwise.

\end{document}